\documentclass[a4paper,12pt]{svjour3_arxiv}
\usepackage{makeidx}
\usepackage{amsfonts}
\usepackage{amsmath}
\usepackage{amssymb}
\usepackage{a4wide}
\usepackage[dvips]{graphicx}
\usepackage{algorithmic}
\usepackage{algorithm}
\usepackage[bottom]{footmisc}
\usepackage{enumerate}

\spnewtheorem{observation}[problem]{Observation}{\bfseries}{\itshape}
\newcommand{\heading}[1]{\medskip\par\noindent{\bf #1}}

\newenvironment{packed_itemize}{
	\begin{itemize}
		\setlength{\itemsep}{1pt}
	    \setlength{\parskip}{0pt}
		\setlength{\parsep}{0pt}
}{\end{itemize}}

\def\O{\mathcal{O}{}}
\def\blt{\blacktriangleleft}

\def\wlt{\vartriangleleft}

\def\computationproblem#1#2#3{
	\begin{center}
	\begin{tabular}{rp{10cm}}
	{\bf Problem:\enspace}&#1\\
	{\bf Input:\enspace}&#2\\
	{\bf Output:\enspace}&#3\\
	\end{tabular}
	\end{center}
}

\def\recog{\textsc{Recog}}
\def\ext{\textsc{RepExt}}

\def\partition{\textsc{3-Partition}}
\def\binpack{\textsc{BinPacking}}
\def\genbinpack{\textsc{GenBinPacking}}
\def\fixed{\textsc{Fixed}}
\def\sub{\textsc{Sub}}
\def\add{\textsc{Add}}
\def\both{\textsc{Both}}
\def\int{\hbox{\bf \rm \sffamily INT}}
\def\pint{\hbox{\bf \rm \sffamily PINT}}

\def\path{\hbox{\bf \rm \sffamily PATH}}
\def\chor{\hbox{\bf \rm \sffamily CHOR}}

\def\boldint{\hbox{\bf \sffamily INT}}
\def\boldpint{\hbox{\bf \sffamily PINT}}

\def\calP{{\cal P}}
\def\calC{{\cal C}}
\def\calR{{\cal R}}
\def\frakT{\mathfrak{T}}
\def\minspan{{\rm minspan}}
\def\wtt{{\rm wtt}}
\def\cl{{\rm cl}}
\def\gen{{\rm Gen}}

\def\cP{\hbox{\rm \sffamily P}}
\def\cNP{\hbox{\rm \sffamily NP}}
\def\cW#1{\hbox{\rm \sffamily W[#1]}}
\def\cXP{\hbox{\rm \sffamily XP}}
\def\cFPT{\hbox{\rm \sffamily FPT}}

\begin{document}

\title{Extending Partial Representations of Subclasses\\of Chordal Graphs}
\newcounter{lth}
\setcounter{lth}{1}
\author{Pavel Klav\'{\i}k \and Jan Kratochv\'{\i}l \and Yota Otachi \and Toshiki Saitoh}
\institute{
		Pavel Klav\'{\i}k \at
		Computer Science Institute, Faculty of Mathematics and
   		Physics, Charles University in Prague, Malostransk{\'e} n{\'a}m{\v e}st{\'\i} 25,
        118 00 Prague, Czech Republic. \email{klavik@iuuk.mff.cuni.cz}
		\and 
		Jan Kratochv\'{\i}l \at
		Department of Applied Mathematics, Faculty of Mathematics and
	   	Physics, Charles University in Prague, Malostransk{\'e} n{\'a}m{\v e}st{\'\i} 25,
        118 00 Prague, Czech Republic. \email{honza@kam.mff.cuni.cz}
		\and 
		Yota Otachi \at
		School of Information Science, Japan Advanced Institute of
		Science and Technology. Asahidai 1-1, Nomi, Ishikawa 923-1292, Japan.
		\email{otachi@jaist.ac.jp}
		\and 
		Toshiki Saitoh \at
		Graduate School of Engineering, Kobe University, Rokkodai 1-1, Nada, Kobe, 657-8501, Japan.
		\email{saitoh@eedept.kobe-u.ac.jp}
}
\authorrunning{P.~Klav\'\i k, J.~Kratochv\'{\i}l, Y.~Otachi, T.~Saitoh}

\maketitle

\begin{abstract}
Chordal graphs are intersection graphs of subtrees of a tree $T$. We investigate the complexity of
the partial representation extension problem for chordal graphs. A partial representation specifies
a tree $T'$ and some pre-drawn subtrees of $T'$. It asks whether it is possible to construct a
representation inside a modified tree $T$ which extends the partial representation (i.e, keeps the
pre-drawn subtrees unchanged).

We consider four modifications of $T'$ and get vastly different problems.  In some cases, it is
interesting to consider the complexity even if just $T'$ is given and no subtree is pre-drawn. Also, we
consider three well-known subclasses of chordal graphs: Proper interval graphs, interval graphs and
path graphs.  We give an almost complete complexity characterization.

We further study the parametrized complexity of the problems when parametrized by the number of
pre-drawn subtrees, the number of components and the size of the tree $T'$. We describe an
interesting relation with integer partition problems.  The problem \partition\ is used for all
\cNP-completeness reductions.  The extension of interval graphs when the space in $T'$ is limited is
``equivalent'' to the \binpack\ problem.
\end{abstract}

\section{Introduction}

Geometric representations of graphs and graph drawing are important topics of graph theory. We
study \emph{intersection representations} of graphs where the goal is to assign geometrical objects
to the vertices of the graph and encode edges by intersections of these objects. An
intersection-defined class restricts the geometrical objects and contains all graphs representable
by these restricted objects; for example, interval graphs are intersection graphs of closed
intervals of the real line. Intersection-defined classes have many interesting properties and appear
naturally in numerous applications; for details see for example~\cite{agt,egr,tig}.

For a fixed class, its recognition problem asks whether an input graph belongs to this class; in
other words, whether it has an intersection representation of this class. The complexity of
recognition is well-understood for many classes; for example interval graphs can be recognized in
linear-time~\cite{PQ_trees,LBFS_int}.

We study a recently introduced generalization of the recognition problem called the \emph{partial
representation extension}~\cite{kkv}. Given a graph and a partial representation (a representation
of an induced subgraph), it asks whether it is possible to extend this partial representation to a
representation of the entire graph. This problems falls into the paradigm of extending partial
solutions, an approach that has been studied frequently in other circumstances. Often it proves to
be much harder than building a solution from scratch, for example for graph
coloring~\cite{prext_bip,fiala}.  Surprisingly, a very natural problem of extending partially
represented graphs was only considered recently.

The paper~\cite{kkv} gives an $\O(n^2)$-algorithm for interval graphs and an $\O(nm)$-algorithm for
proper interval graphs.  Also, several other papers consider this problem. Interval representations
can be extended in time $\O(n+m)$~\cite{blas_rutter,kkosv}. Proper interval representations can be
extended in time $\O(n+m)$ and unit interval representations in time $\O(n^2)$~\cite{kkorssv}.
Polynomial time algorithms are also described for function and permutation graphs~\cite{kkkw}, and for
circle graphs~\cite{cfk}.

In this paper, we follow this recent trend and investigate the complexity of partial representation
extension of chordal graphs. Our mostly negative \cNP-completeness results are very interesting
since chordal graphs are the first class for which the partial representation problem is proved to
be strictly harder than the original recognition problem. Also, we investigate three well-known
subclasses -- proper interval graphs, interval graphs and path graphs, for which the complexity results
are richer. We believe that better understanding of these simpler cases will provide tools to attack
chordal graphs and beyond (for example, from the point of the parameterized complexity). For the
conference version of this paper see~\cite{kkos}.

\subsection{Chordal Graphs and Their Subclasses}

A graph is \emph{chordal} if it does not contain an induced cycle of length four or
more, i.e., each ``long'' cycle is triangulated.  The class of chordal graphs, denoted by \chor, is
well-studied and has many wonderful properties. Chordal graphs are closed under induced subgraphs
and possess the so called \emph{perfect elimination schemes} which describe perfect reorderings of
sparse matrices for the Gaussian elimination. Chordal graphs are perfect and many hard combinatorial
problems are easy to solve on chordal graphs: maximum clique, maximum independent set, $k$-coloring,
etc.  Chordal graphs can be recognized in time $\O(n+m)$~\cite{recog_chordal_graphs}.

\begin{figure}[t]
\centering
\includegraphics[scale=0.975]{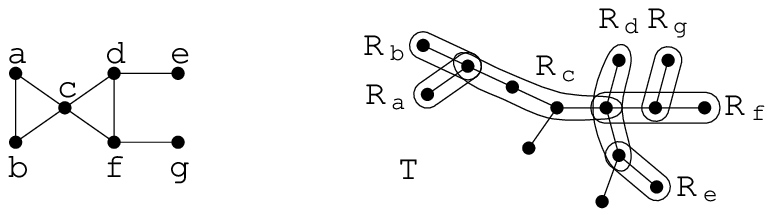}
\caption{An example of a chordal graph with one of its representations.}
\label{fig:chordal_example}
\end{figure}

Chordal graphs have the following intersection representations~\cite{chor_is_subtree_in_tree}. For
every chordal graph $G$ there exists a tree $T$ and a collection $\{R_v \mid v \in V(G)\}$ of subtrees
of $T$ such that $R_u \cap R_v \ne \emptyset$ if and only if $uv \in E(G)$.  For an example of a
chordal graph and one of its intersection representations, see Fig.~\ref{fig:chordal_example}.

When chordal graphs are viewed as \emph{subtrees-in-tree} graphs, it is natural to consider two
other possibilities: \emph{subpaths-in-path} which gives \emph{interval graphs} (\int), and
\emph{subpaths-in-tree} which gives \emph{path graphs} (\path). For example the graph in
Fig.~\ref{fig:chordal_example} is a path graph but not an interval one. Subpaths-in-path
representations of interval graphs can be viewed as discretizations of the real line
representations. Interval graphs can be recognized in $\O(n+m)$~\cite{PQ_trees,LBFS_int} and path
graphs in time $\O(nm)$~\cite{path_original,path_fastest}.

In addition, we consider proper interval graphs ($\pint$). An interval graph is a \emph{proper interval
graph} if it has a representation $\calR$ for which $R_u \subseteq R_v$ implies $R_u = R_v$;
so no interval is a proper subset of another one.\!\footnote{%
It is possible to define proper interval graphs differently: If $R_u \subseteq R_v$, then $R_v
\setminus R_u$ is empty or a connected subpath of $T$. In other words, no interval can be placed in
the middle of another interval. Our results can be easily modified for this alternative definition.}
Proper interval graphs can be recognized in time~$\O(n+m)$~\cite{proper_first,uint_corneil}.  From
the point of our results, \pint\ behaves very similar to \int\ but there are subtle differences
which we consider interesting. Also, partial representation extension of \pint\ is surprisingly very
closely related to partial representation extension of unit interval graphs considered
in~\cite{kkorssv}; see Section~\ref{sec:intro_related} for details.

\subsection{Partial Representation Extension}

For a class $\calC$, we denote the recognition problem by $\recog(\calC)$. For an input graph $G$,
it asks whether it belongs to $\calC$, and moreover we may certify it by a representation.  The
partial representation extension problem denoted by $\ext(\calC)$ asks whether a part of the
representation given by the input can be extended to a representation of the whole graph.

A partial representation $\calR'$ of $G$ is a representation of an induced subgraph $G'$.
The vertices of $G'$ are called \emph{pre-drawn}. A representation $\calR$ extends $\calR'$ if $R_v
= R'_v$ for every $v \in V(G')$.  The meta-problem we deal with is the following. 

\computationproblem
{$\ext(\calC)$ (Partial Representation Extension of $\calC$)}
{A graph $G$ with a partial representation $\calR'$.}
{Does $G$ have a representation $\calR$ that extends $\calR'$?}

\begin{figure}[t]
\centering
\includegraphics[scale=0.975]{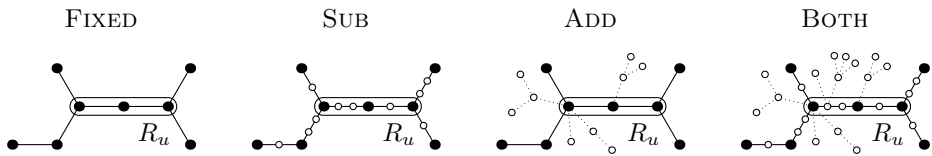}
\caption{The four possible modifications of $T'$ with a single pre-drawn vertex $u$. The added
branches in $T$ are denoted by dots and new vertices of $T$ are denoted by small circles.}
\label{fig:tree_modifications}
\end{figure}

In this paper, we study complexity of the partial representation extension problems for the classes \chor,
\path, \int, and \pint\ in the setting of subtrees-in-tree representations. Here a partial
representation $\cal R'$ fixes subtrees belonging to $G'$ and also specifies some tree $T'$ in which
these subtrees are placed. A representation $\calR$ is placed in a tree $T$ which is created by
some modification of $T'$. We consider four possible modifications and get different extension
problems:
\begin{packed_itemize}
\item \fixed\ -- the tree cannot be modified at all, i.e, $T = T'$.
\item \sub\ -- the tree can only be subdivided, i.e., $T$ is a subdivision of $T'$.\footnote{Let an edge
$xy \in E(T')$ be subdivided (with a vertex $z$ added in the middle). Then also pre-drawn subtrees
containing both $x$ and $y$ are modified and contain $z$ as well. So technically in the case of
subdivision, it is not true that $R'_u = R_u$ for every pre-drawn interval, but from the topological
point of view the partial representation is extended.}
\item \add\ -- we can add branches to the tree, i.e., $T'$ is a subgraph of $T$.
\item \both\ -- we can both add branches and subdivide, i.e, a subgraph of $T$ is a subdivision of
$T'$. In other words $T'$ is a topological minor of $T$.
\end{packed_itemize}
We denote the problems by $\ext(\calC,\frakT)$ where $\frakT$ denotes the type. See
Fig.~\ref{fig:tree_modifications}.

Constructing a representation in a specified tree $T'$ is interesting even if no subtree is
pre-drawn, i.e., $G'$ is empty; this problem is denoted by $\recog^*(\calC,\frakT)$. Clearly,
the hardness of the $\recog^*$ problem implies the hardness of the corresponding $\ext$ problem.

For \pint\ and \int\ classes, the types \add\ and \sub\ behave as follows. The type \add\ allows
to extend the ends of the paths. The type \sub\ allows to expand the middle of the path. The
difference is that if an endpoint of the path is contained in some pre-drawn subpath, it remains
contained in it after the subdivision. The type \both\ makes the problems equivalent to the \recog\
and \ext\ problems for the real line.

\subsection{Our Results}

\begin{figure}[t!]
\centering
\includegraphics{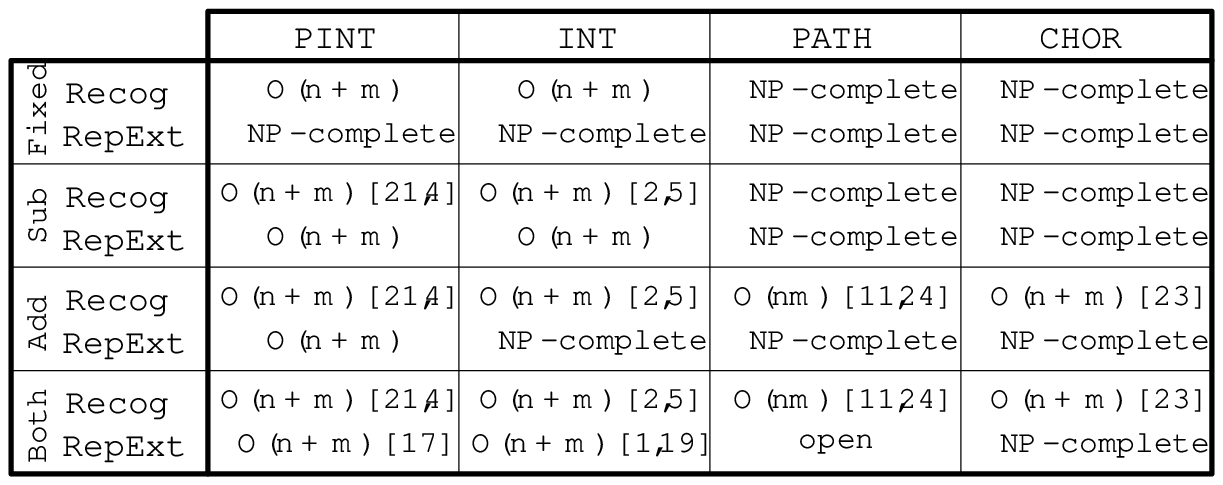}
\caption{The table of the complexity of different problems for the four considered classes. The
results without references are new results of this paper.}
\label{fig:complexity_table}
\end{figure}

We study the complexity of the $\recog^*$ and \ext\ problems for all four classes and all four
types. Our results are displayed in Fig.~\ref{fig:complexity_table}.
\begin{packed_itemize}
\item All \cNP-complete results are reduced from the \partition\ problem. The reductions are very
similar and the basic case is Theorem~\ref{thm:int_npc} for $\ext(\int,\fixed)$ and
$\ext(\pint,\fixed)$.
\item The polynomial cases for \int\ and \pint\ are based on the known algorithm for
recognition and extension. But since the space in $T$ is limited, we adapt the algorithm for the
specific problems.
\end{packed_itemize}

Every interval graph has a real-line representation in which all endpoints are at integer positions.
But the result that $\ext(\int,\add)$ is \cNP-complete can be interpreted in the way that extending
such representations is \cNP-complete. (Here, we require that also the non-pre-drawn intervals have
endpoints placed at integer positions.) On the other hand, our linear-time algorithm for
$\ext(\pint,\add)$ shows that integer-position proper interval representations can be extended in
linear time.

For a subpaths-in-path partial representation, we assume that an input gives the endpoints of the
pre-drawn subpaths sorted by the input from left to right. This allows us to construct algorithms in
time $\O(n+m)$ which do not depend on the size of the path $T'$.

\heading{Parameterized Complexity.}
We study the parameterized complexity of these problems with respect to three parameters: The number
$k$ of pre-drawn subtrees, the number $c$ of components and the size $t$ of the tree $T'$. In some
cases, the parametrization does not help and the problem is \cNP-complete even if the value of the
parameter is zero or one. In other cases, the problems are fixed-parameter tractable (\cFPT),
\cW1-hard or in \cXP.

The main result concerning parametrization is the following. The \binpack\ problem is a well-known
problem concerning integer partitions; more details in Section~\ref{sec:int_param}.  For two
problems $A$ and $B$, we denote by $A \le B$ polynomial reducibility and by $A \le_{\wtt} B$ weak
truth-table reducibility. (Roughly speaking, to solve $A$ we may use a number of $B$-oraculum
questions which is bounded by a computable function.)

\begin{theorem} \label{thm:pint_equiv_bin_pack}
For the number $k$ of bins and pre-drawn subtrees, we get
$$\binpack \le \ext(\pint,\fixed) \le_{\wtt} \binpack.$$
The weak truth-table reduction needs to solve $2^k$ instances of \binpack.
\end{theorem}

\subsection{Two Related Problems} \label{sec:intro_related}

We describe two problems which are closely related to our results.

The problem of \emph{simultaneous representations}~\cite{simrep_chor} asks whether there exist
representations $\calR_1,\dots,\calR_k$ of graphs $G_1,\dots,G_k$ which are the same on the common
part of the vertex set $I = V(G_i) \cap V(G_j)$ for all $i \ne j$. It is noted in~\cite{kkv} that
the partial representation extension is closely related to the simultaneous representations. For
instance, using simultaneous representations of interval graphs, we can solve their partial
representation extension~\cite{blas_rutter}. As we show in this paper, this is not the case for
chordal graphs since $\ext$ of chordal graphs is \cNP-complete but their simultaneous
representations are solvable in polynomial-time~\cite{simrep_chor}.

The partial representation extension problem of proper interval graphs described here is closely
related to partial representations and the \emph{bounded representation problem} of unit interval
graphs~\cite{kkorssv}. In all these problems, one deals with interval representations in a limited
space. So the techniques initially developed for unit interval graphs are easily used here for
proper interval graphs. We note that the problems concerning unit interval graphs are more difficult
since they involve computations with rational number positions.

\section{Preliminaries} \label{sec:prelim}

In this section, we describe the notation used in this paper. Also, we deal with two common concepts
of the partial representation extension problems: Located and unlocated components, and groups of
indistinguishable vertices.

\heading{Notation.}
We consider finite undirected simple graphs, i.e., graphs without loops and multiedges. As usual,
we reserve $n$ for the number of the vertices and $m$ for the number of the edges of the main
considered graph $G$. The set of its vertices is denoted by $V(G)$ and the set of its edges by
$E(G)$. For a vertex $v \in V(G)$, we let $N(v)=\{x \mid vx\in E(G)\}$ denote the \emph{open
neighborhood} of $v$, and $N[v] = N(v)\cup\{v\}$ the \emph{closed neighborhood} of $v$.

By $P_n$, we denote the path of the length $n$ with $n+1$ vertices. For a tree, we call the vertices
of degree larger than two \emph{branch vertices} and the vertices of degree at most two
\emph{non-branch vertices}, and of course the vertices of degree one are called \emph{leaves}.

\begin{figure}[b!]
\centering
\includegraphics[scale=0.85]{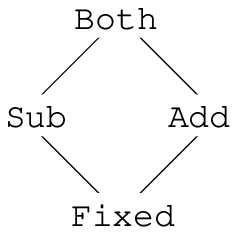}
\caption{The lattice formed by four types \fixed, \sub, \add, and \both.}
\label{fig:type_lattice}
\end{figure}

\heading{The Type Lattice.}
The four types \fixed, \sub, \add, and \both\ form the lattice depicted in
Fig.~\ref{fig:type_lattice}. For a type $\frakT$, we denote by $\gen(\frakT,T')$ the set of all
trees $T$ which we can generate from $T'$ using the modifications of the type $\frakT$. In addition,
if $T'$ contains pre-drawn subtrees, the trees in $\gen(\frakT,T')$ contain these (possibly
subdivided) pre-drawn subtrees as well. The ordering of the types given by the lattice has this
property: If $\frakT \le \frakT'$, then $\gen(\frakT,T') \subseteq \gen(\frakT',T')$.

Whether a given instance is solvable depends on the set $\gen(\frakT,T')$; so if this set contains
more trees, it only helps in solving the problem. Let $\frakT \le \frakT'$. If an instance of
$\recog^*$ or \ext\ is solvable for the type $\frakT$, then it is solvable for the type $\frakT'$ as
well.  Equivalently, if it is not solvable for $\frakT'$, it is also not solvable for $\frakT$.

For the types \add\ and \both\ (and \sub\ for \pint\ and \int), the set $\gen(\frakT,T')$ contains a
tree having an arbitrary tree $T$ as a subtree. Therefore, the $\recog^*$ problem for these types is
equivalent to the standard $\recog$ problem, and we can use the known polynomial-time algorithms.

\heading{Topology of Components.} The following property works quite generally for many
intersection-defined classes of graphs, and works for all classes studied in this paper. The only
required condition is that the sets $R_v$ are connected subsets of some topological space, for
example $\mathbb R^k$.  (As a negative example, this property does not hold for \emph{$2$-interval
graphs}.  A graph is a $2$-interval graph if each $R_v$ is a union of two closed intervals.) Let $C$
be a connected component of $G$. Then the property is that for each representation $\calR$, the set
$\bigcup_{v \in C} R_v$ is a connected subset of the space, and we call this subset the \emph{area of
$C$}. Clearly, the areas of the components are pairwise disjoint.

For the classes \pint\ and \int, the areas of the components have to be ordered from left to right.
Let us denote this ordering by $\blt$, so we have $C_1 \blt \cdots \blt C_c$. For different
representations $\calR$, we can have different orderings $\blt$. When no restriction is posed on
$\calR$, it is possible to create a representation in every of the $c!$ possible orderings.

\heading{Types of Components.}
For the partial representation extension problem, the graph $G$ contains two types of components.  A
component $C$ is called a \emph{located component} if it has at least one vertex pre-drawn, i.e., $C
\cap G'$ is non-empty. A component $C$ is called an \emph{unlocated component} if no interval is
pre-drawn, i.e., $C \cap G' = \emptyset$. For located components, we have a partial information about
their position.  For unlocated components, we are much freer in their placement.

For the classes of interval graphs, the located components are ordered from left to right. An
obvious necessary condition for an extendible partial representation is that the pre-drawn intervals
of each component appear consecutively in $\calR'$. Indeed, if $C$ and $C'$ are two distinct
components, $u,v \in C$, $w \in C'$ and $R_w$ is between $R_u$ and $R_v$, then the partial
representation is clearly not extendible.  For every representation $\calR$ extending $\calR'$, the
ordering $\blt$ has to extend the ordering $\blt'$ of the located components in $\calR'$.

For many of the considered problems the unlocated components are irrelevant. For instance for
$\ext(\int,\both)$, we can extend the path $T$ far enough to the right and place the unlocated
components there, without interfering with the partial representation at all. On the other hand, for
problems involving the types \fixed\ and \sub, the space in $T$ is limited and the unlocated
components have to be placed somewhere. In many cases, the existence of unlocated components is not
only used for \cNP-completeness proofs but also necessary for the problems to be \cNP-complete.

\heading{Indistinguishable Vertices.}
Let $u$ and $v$ be two vertices of $G$ such that $N[u] = N[v]$. These two vertices are called
\emph{indistinguishable} since they can be represented exactly the same, i.e., $R_u = R_v$. (This
is a common property of indistinguishable vertices for all intersection representations). From the
structural point of view, \emph{groups of indistinguishable} vertices are not very interesting. The
goal is to construct a pruned graph where each group is represented by a single vertex. For that, we
need to be little careful since we cannot prune pre-drawn vertices.

For an arbitrary graph, its groups of indistinguishable vertices can be located in time
$\O(n+m)$~\cite{recog_chordal_graphs}. We prune the graph in the following way. If $u$ and $v$ are
indistinguishable and $u$ is not pre-drawn, we eliminate $u$ from the graph (and for the
representation, we can put $R_u = R_v$). In addition, if two pre-drawn intervals are the same, we
eliminate one of them. The resulting pruned graph has the following property: If
two vertices $u$ and $v$ indistinguishable, they are both pre-drawn and represented by distinct
intervals. For the rest of the paper, we expect that all input graphs are pruned.

\heading{Maximal Cliques.}
It is well known that subtrees of a tree possess the Helly property, i.e.,  every pairwise
intersecting collection of subtrees has a non-empty intersection (which is again a subtree).  Hence
the following holds true for all classes of graphs considered. If $K$ is a maximal clique of $G$,
the common intersection $R_K = \cap_{u \in K} R_u$ is a subtree of $T$. This subtree $R_K$ is not
intersected by any other $R_v$ for $v \notin K$ (otherwise $K$ would not be a maximal clique). Thus
the subtrees $R_K$ corresponding to different maximal cliques are pairwise disjoint. For example, if
$|T|$ is smaller than the number of maximal cliques of $G$, the graph is clearly not representable
in $T$.

\section{Interval Graphs} \label{sec:int}

In this section, we deal with the classes \pint\ and \int. The results obtained here are used as tools
for \path\ and \chor\ in Section~\ref{sec:chor}.

Let $p_1,\dots,p_t$ be the vertices of the path $T'$. For a located component $C$, we say that a
vertex $p_i$ is \emph{taken} by $C$ if there exists a pre-drawn subpath of $C$ containing $p_i$.

\subsection{Structural Results}

We describe two types orderings: Endpoint orderings for proper interval graphs and clique orderings
for interval graphs. Also, we introduce an important concept called the minimum span of a component.

\heading{Endpoint Orderings of \boldpint.}
Each proper interval representation gives some ordering $\wlt$ of the intervals from left to
right. This is the ordering of the left endpoints from left to right, and at the same time the
ordering of the the right endpoints. The following lemma of~\cite{deng} states that $\wlt$ is well
determined:

\begin{lemma}[Deng et al.] \label{pint_ordering}
For a component of a proper interval graph, the ordering $\wlt$ is uniquely determined up to a local
reordering of the groups of indistinguishable vertices and the complete reversal.
\end{lemma}

So for a connected graph, we have a partial ordering $<$ in which exactly the indistinguishable
vertices are incomparable, and each $\wlt$ is a linear extension of either $<$, or its reversal.
Corneil et al.~\cite{uint_corneil} describes how this ordering can be constructed in time $\O(n+m)$.
Since the graphs we consider are pruned, all incomparable vertices in $<$ are ordered by their
positions in the partial representation. Thus we have at most two possibilities for $\wlt$ for each
component $C$.  (And two possibilities only if all pre-drawn vertices are indistinguishable.)

\heading{Minimum Spans of \boldpint.}
For the types \fixed\ and \add, the space on the path $T'$ is limited. So it is important to
minimize the space taken by each component $C$. We call the minimum space required by $C$
the \emph{minimum span of $C$}, denoted by $\minspan(C)$. Let $\calR$ be a proper interval
representation of $C$ extending $\calR'$, and let $p_i$ be the left-most vertex of $T'$ taken by $C$
and $p_j$ the right-most one. Then
$$\minspan(C) = \begin{cases}
\min_{\forall \cal R}\{j-i+1\}&\text{if some representation of $C$ exists,}\\
+\infty&\text{otherwise.}
\end{cases}$$
A representation of $C$ is called \emph{smallest} if it realizes the minimum span of $C$.

\begin{lemma} \label{lem:pint_minspan}
For every component $C$, the value $\minspan(C)$ can be computed in time $\O(n+m)$, together with a
smallest representation of $C$.
\end{lemma}

\begin{proof}
First, we deal with unlocated components, and later modify the approach for the located ones.

\emph{Case 1: An Unlocated Component.} Since there are no indistinguishable vertices, we compute in
time $\O(n+m)$ using the algorithm of~\cite{uint_corneil} any ordering $\wlt$ for which we want to
produce a representation as small as possible.

Let $\ell_i$ denote the left endpoint and $r_i$ the right endpoint of the interval $v_i$.  From the
ordering $v_1 \wlt \cdots \wlt v_n$, we want to compute the common ordering $\lessdot$ of both the
left and the right endpoints from left to right. The starting point is the ordering of just the left
endpoints $\ell_1 \lessdot \cdots \lessdot \ell_n$. Into this ordering, we insert the right endpoints
$r_1,\dots,r_n$ one-by-one. A right endpoint $r_i$ is inserted right before $\ell_j$ where $v_j$ is
the left-most non-neighbor of $v_i$ on the right in $\wlt$; if such $v_j$ does not exist, we append
$r_i$ to the end. For an example of $\lessdot$, see Fig.~\ref{fig:pint_common_ordering}.

\begin{figure}[b]
\centering
\includegraphics{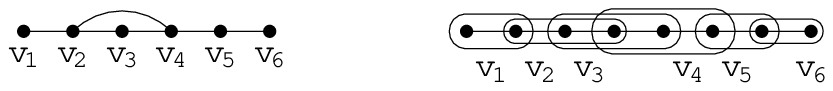}
\caption{The ordering $\lessdot$ is $\ell_1 \lessdot \ell_2 \lessdot r_1 \lessdot \ell_3 \lessdot
\ell_4 \lessdot r_2 \lessdot r_3 \lessdot \ell_5 \lessdot r_4 \lessdot \ell_6 \lessdot r_5 \lessdot
r_6$ for the component $C$ on the left. The constructed smallest possible representation of the
component $C$ on the right, with $\minspan(C) = 8$.}
\label{fig:pint_common_ordering}
\end{figure}

We build a smallest representation using $\lessdot$ as follows. Let $p_1,\dots,p_k$ be the
vertices of the tree $T$. We construct an assignment $f$ which maps the endpoints of the intervals
of $C$ into $T$. Then for a vertex $v_i$ we put
$$R_{v_i} = \{p_j \mid f(\ell_i) \le p_j \le f(r_i)\}.$$
The mapping $f$ is constructed for the endpoints one-by-one, according to $\lessdot$. Suppose that
the previous endpoint in $\lessdot$ has assigned a vertex $p_i$. If the current endpoint is a right
endpoint and the previous endpoint is a left endpoint, we assign $p_i$ to the current endpoint.
Otherwise we assign $p_{i+1}$ to it. For an example, see Fig.~\ref{fig:pint_common_ordering}.

In total, the component needs $2n-\ell$ vertices of $T$ where $\ell$ denotes the number of changes
from a left endpoint to a right endpoint in the ordering $\lessdot$; in other words, $2n-\ell$ is
the value of $\minspan(C)$. The total complexity of the algorithm is clearly $\O(n+m)$.

To conclude the proof, we need to show that we construct a correct smallest representation of $C$.
A property of $\wlt$ is that the closed neighborhood $N[v]$ of every vertex $v \in V(G)$ is
consecutive in $\wlt$.  If $v_iv_j \in E(G)$ and $v_i \wlt v_j$, then $\ell_i \lessdot \ell_j
\lessdot r_i$, and so $R_{v_i}$ intersects $R_{v_j}$ (between $f(\ell_j)$ and $f(r_i)$). If $v_iv_j
\notin E(G)$, then $r_i \lessdot \ell_j$. Thus $r_i$ is placed on the left of $\ell_j$ in
$\lessdot$, and $R_{v_i} \cap R_{v_j} = \emptyset$ as required.

Concerning the minimality notice that in a pruned graph, $\ell_i \ne \ell_j$ and $r_i \ne r_j$ hold
for every $i \ne j$. We argue that we use gaps as small as possible. Only a right endpoint $r_i$
following a left endpoint $\ell_j$ can be placed at the same position. The other case of a right
endpoint $r_i$ followed by a left endpoint $\ell_j$ requires a gap of size one; otherwise
$R_{v_i}$ would intersect $R_{v_j}$ but $v_iv_j \notin E(G)$. So the gaps are minimal, we
construct a smallest representation, and give the value $\minspan(C)$ correctly.

\emph{Case 2: A Located Component.} We modify the above approach slightly to deal with located
components. We already argued that there are at most two possible orderings $\wlt$ (since
the indistinguishable vertices are ordered by the partial representation), and we just test both of them.
Both orderings can be used if and only if all pre-drawn vertices belong to one group of
indistinguishable vertices. Then these two orderings give the same $\minspan(C)$ but the minimum
representations might be differently shifted, and we are able to construct both of them. If the
pre-drawn intervals do not belong to one group, the ordering $\wlt$ is uniquely determined. (If it is
compatible with the ordering of the pre-drawn intervals at all.)

We compute the common ordering $\lessdot$ exactly as before and place the endpoints in this
ordering. The only difference is that the endpoints of the pre-drawn intervals are prescribed.
So we start at the position of the left-most pre-drawn endpoint $\ell_i$. We place the endpoints
smaller in $\lessdot$ than $\ell_i$ on the left of $\ell_i$ as far to the right as possible. (We
approach them in the reverse order exactly as above.) Then we proceed with the remaining endpoints in the
order given by $\lessdot$. If the current endpoint is pre-drawn, we keep it as it is. Otherwise, we
place it in the same way as above. The constructed representation is smallest and gives
$\minspan(C)$.\qed
\end{proof}

\heading{Clique Orderings of \boldint.}
Recall the properties of maximal cliques from Section~\ref{sec:prelim}. For a component $C$, we denote
by $\cl(C)$ the number of maximal cliques of $C$. Let $\calR$ be a representation of $C$. Since the
subtrees $R_K$ corresponding to the maximal cliques are pairwise disjoint, they have to be ordered
from left to right. This ordering has the following well-known property~\cite{maximal_cliques}:

\begin{lemma}[Fulkerson and Gross] \label{interval_char}
A graph is an interval graph if and only if there exists an ordering of the maximal cliques $K_1 <
\cdots < K_{\cl(C)}$ such that for each vertex $v$ the cliques containing $v$ appear consecutively
in this ordering.
\end{lemma}

We quickly argue about the correctness of the lemma.  Clearly, in an interval representation, all
maximal cliques containing one vertex $v$ appear consecutively. (Otherwise the clique in between
would be intersected by $R_v$ in addition.) On the other hand, having an ordering $<$ of the maximal
cliques from the statement, we can construct a representation as follows.  Assign a vertex $p_i$ of
$T$ to each clique $K_i$, respecting the ordering $<$.  For each vertex $v$, we assign $R_v = \{p_i
\mid v \in K_i\}$.  Since the maximal cliques containing $v$ appear consecutively, each $R_v$ is a
subpath.

\heading{Minimum Spans of \boldint.}
We again consider the minimum span defined exactly as for proper interval
graphs above. Clearly, $\minspan(C) \ge \cl(C)$. We show:

\begin{lemma} \label{lem:int_minspan}
For an unlocated component $C$ of an interval graph, $\minspan(C) = \cl(C)$.  We can find a smallest
representation in time $\O(n+m)$.
\end{lemma}

\begin{proof}
We start by identifying maximal cliques in time $\O(n+m)$, using the algorithm of Rose et
al.~\cite{recog_chordal_graphs}. To construct a smallest representation, we find an ordering from
Lemma~\ref{interval_char}, using the PQ-tree algorithm~\cite{PQ_trees} in time $\O(n+m)$. If such
an ordering does not exist, the graph $G$ is not an interval graph and no representation exists. If the
ordering exists, we can construct a representation using exactly $\cl(C)$ vertices of the path as
described above, by putting $R_v = \{p_i \mid v \in K_i\}$.\qed
\end{proof}

We note that this approach does not translate to located components, as in
Lemma~\ref{lem:pint_minspan} for proper interval graphs. We prove in Corollary~\ref{cor:int_add_npc}
that finding the minimum span for a located component is an \cNP-complete problem. (We prove this in the
setting that the problem $\ext(\int,\add)$ is \cNP-complete. In the reduction, we ask whether a
connected interval graph has the minimum span at most $(M+1)k+1$ for some integers $k$ and $M$.)

\subsection{The Polynomial Cases}

First we deal with all polynomial cases.

\heading{Fixed Type Recognition.} We just need to use the values of minimum spans we already know
how to compute.

\begin{proposition} \label{prop:int_fixed}
Both $\recog^*(\pint,\fixed)$ and $\recog^*(\int,\fixed)$ can be solved in time $\O(n+m)$.
\end{proposition}

\begin{proof}
We process the components $C_1,\dots,C_c$ one-by-one and place them on $T'$ from left to right. If
$\sum_{i=1}^c \minspan(C_i) \le |T'|$, we can place the components using smallest representations
from Lemma~\ref{lem:pint_minspan} for \pint, resp.~Lemma~\ref{lem:int_minspan} for \int. Otherwise,
the path is too small and a representation cannot be constructed.\qed
\end{proof}

\heading{Add Type Extension, \boldpint.}
Again, we approach this problem using minimum spans and Lemma~\ref{lem:pint_minspan}.

\begin{proposition} \label{prop:pint_add}
The problem $\ext(\pint,\add)$ can be solved in time $\O(n+m)$.
\end{proposition}

\begin{proof}
Since the path can be expanded to the left and to the right as much as necessary, we can place unlocated
components far to the left. So we only need to deal with located components, ordered $C_1 \blt
\cdots \blt C_c$ from left to right. We process the components from left to right. When we place
$C_i$, it has to be placed on the right of $C_{i-1}$. We have (at most) two possible smallest representations
corresponding to two different orderings of $C_i$. We test whether at least one of them can be
placed on the right of $C_{i-1}$, and pick the one minimizing the right-most vertex of $T$ taken by
$C_i$ (leaving the maximum possible space for $C_{i+1},\dots,C_c$). If neither of the smallest
representations can be placed, the extension algorithm outputs ``no''.

If the algorithm finishes, it constructs a correct representation. On the other hand, we place each
component as far to the left as possible (while restricted by the previous components on the left).
So if $C_i$ cannot be placed, there exists no representation extending the partial
representation.\qed
\end{proof}

\heading{Non-fixed Type Recognition.}
The only limitation for recognition of interval graphs inside a given path is the length of the
path. In the three types \sub, \add\ and \both, we can produce a path as long as necessary. (With
the trivial exception $T' = P_0$ for \sub\ for which the instance is solvable if and only if $G=K_n$.)
For a subpaths-in-path representation, the order of the endpoints of the subpaths from left to right
is the only thing that matters, not the exact positions. In a tree $T$ with at least $2n$ vertices,
every possible ordering is realizable.

Thus the problems are equivalent to the standard recognition of interval graphs on the real line.
The recognition can be solved in time $\O(n+m)$; see~\cite{proper_first,uint_corneil} for \pint,
and~\cite{PQ_trees,LBFS_int} for \int.

\heading{Both Type Extension.}
This extension type is equivalent with the partial representation extension problems of interval
graphs on the real line. Again only the ordering of the endpoints is important. The only change here
is that some of the endpoints are already placed. By subdividing, we can place any amount of the
endpoints between any two endpoints (not sharing the same position). Also, the path can be extended
to the left and to the right which allows to place any amount of endpoints to the left of the
left-most pre-drawn endpoint and to the right of the right-most pre-drawn endpoint. So any extending
ordering can be realized in the \both\ type.

The partial representation extension problem for interval graphs on the real line was first
considered in~\cite{kkv}. The paper gives algorithms for both classes \int\ and \pint, and does not
explicitly deal with representations sharing endpoints but the algorithms are easy to modify.
The results~\cite{blas_rutter,kkosv,kkorssv} show that both extension problems are solvable in
time $\O(n+m)$.

\heading{Sub Type Extension.}
It is possible to modify the above algorithms for partial representation extension of \int\ and
\pint. Instead of describing details of these algorithms, we simply reduce the problems to the type
\both\ which we can solve in time $\O(n+m)$ (as discussed above):

\begin{theorem} \label{thm:int_sub}
The problems $\ext(\pint,\sub)$ and $\ext(\int,\sub)$ can be solved in time $\O(n+m)$.
\end{theorem}

The general idea is as follows. The difference between between \sub\ and \both\ is that for the
\sub\ type, we cannot extend the path $T'$ at the ends. Suppose that some pre-drawn subpath $R'_v$
contains say the left endpoint of $T'$. Then $R'_v$ contains this endpoint also in $T$. So we are
going to modify the graph $G$ in such a way, that every representation of the \both\ type has to
place everything on the right of $R'_v$.

Suppose first that the graph contains some unlocated components, and we show how to deal with them.
We want to find one edge $p_ip_{i+1}$ of $T'$ which we can subdivide many times and place all
unlocated components in between of $p_i$ and $p_{i+1}$ in $T$. We call an edge $p_ip_{i+1}$
\emph{expandable} if no located component $C$ takes $p_j$ and $p_k$ such that $j \le i < i+1 \le k$.

\begin{lemma} \label{lem:dealing_unlocated}
Let $G$ have at least one unlocated component, and let $\widetilde G$ be the graph constructed from
$G$ by removing all unlocated components. Then $\calR'$ is extendible to $\calR$ if and only if
\,$T'$\!  contains at least one expandable edge $p_ip_{i+1}$ and $\calR'$ is extendible to
$\widetilde\calR$ of $\widetilde G$.
\end{lemma}

\begin{proof}
Let $\calR'$ be extendible to $\calR$ and let $C$ be one unlocated component placed in $T$ such that
it takes a vertex in between of $p_i$ and $p_{i+1}$ of $T'$. Clearly $\calR'$ is extendible to
$\widetilde\calR$. And $p_ip_{i+1}$ is expandable since if there would be a located component
$\widetilde C$ taking $p_j$ and $p_k$, then $C$ would split $\widetilde C$, contradicting existence
of $\blt$ in $\calR$; recall the definition of $\blt$ in Section~\ref{sec:prelim}.

For the other implication, we subdivide the expandable edge $p_ip_{i+1}$ many times such that we can
place all unlocated components in  this area. For located components, some of them have
to be placed on the left of the unlocated components, and some on the right. We can subdivide all edges
of $T'$ enough to place the endpoints in the same order as in $\widetilde\calR$. Thus we get $\calR$
extending $\calR'$.\qed
\end{proof}

\begin{proof}[Theorem~\ref{thm:int_sub}]
We describe the reduction for \int, and then we slightly modify it in the last paragraph for \pint.
We deal with unlocated components using Lemma~\ref{lem:dealing_unlocated}. We just need to check
existence of an expandable edge for which we first compute the ordering $\blt$ of the located
components (if it doesn't exist, the partial representation is clearly not extendible).  If there is
exactly one located component $C$, then at least one of $p_1$ and $p_t$ is not taken by $C$, and say
for $p_1$ we obtain an expandable edge $p_1p_2$. And if there are at least two located components
$C_1 \blt \cdots \blt C_c$, let $p_i$ be the right-most vertex taken by $C_1$. Then $p_ip_{i+1}$
is clearly expandable. It remains to deal with located components.

Let us consider the endpoint $p_1$ of $T'$. In \both, we can attach in $T$ a path $P$ of any length on the left
of $p_1$. If $p_1$ is not taken by $C_1$, we can create in $T$ the same path $P$ by subdividing $p_1p_2$.
But if $p_1$ is taken by $C_1$, we have to forbid $P$ to be used in the construction of $\calR$.
We modify both the path $T'$ and the graphs $G$, and we show that any representation $\calR$
extending $\calR'$ is realized in $T$ in between of $p_1$ and $p_t$.

The modification is as follows. Let $v_1,\dots,v_k \in C_1$ be all pre-drawn subpaths such that $p_1 \in
R'_{v_1},\dots,R'_{v_k}$. First, we extend the path by one by adding $p_0$ attached to $p_1$. We
introduce an additional pre-drawn vertex $v_\leftarrow$ adjacent exactly to $v_1,\dots,v_k$ in $G$.
We put $R'_{v_\leftarrow} = \{p_0\}$ and we modify $R'_{v_i} = R'_{v_i} \cup \{p_0\}$. See
Fig.~\ref{fig:modified_sub_representation}.
Indeed, we proceed exactly the same on the other side of $T'$; if $p_t$ is taken by $C_c$, we
introduce $p_{t+1}$ and $v_\rightarrow$.

\begin{figure}[t]
\centering
\includegraphics{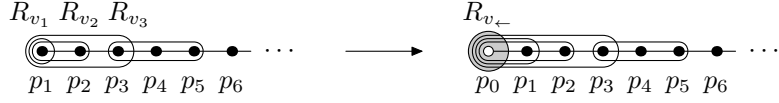}
\caption{The three pre-drawn subpaths containing $p_1$ are $R_{v_1} = \{p_1\}$, $R_{v_2} =
\{p_1,p_2\}$ and $R_{v_3} = \{p_1,p_2,p_3\}$. We add $p_0$ to $T'$ and to $R_{v_1},\dots,R_{v_3}$,
and we introduce additional pre-drawn subpaths $R_{v_\leftarrow} = \{p_0\}$.}
\label{fig:modified_sub_representation}
\end{figure}

We use the described algorithm for $\ext(\int,\both)$ for the modified graph and the modified path,
which runs in time $\O(n+m)$. We obtain a representation $\calR$ extending $\calR'$ if it exists.
If $\calR$ does not exist, then the original problem is clearly not solvable. It remains to argue that if
$\calR$ exists, then we can either construct a solution for the original \sub\ type problem, or we
can prove that it is not solvable.

We deal only with the left side of $T$; for the right side the argument is symmetrical. If $T'$ is
not modified on the left side, then the edge $p_1p_2$ can be subdivided as necessary and we are
equivalent with the \both\ type. Suppose that $p_0$ is added.  There are no unlocated components,
and so everything with the exception of $v_1,\dots,v_k$ has to be represented on the right of
$v_\leftarrow$ which is placed on $p_0$.

We need to argue the issue that the newly added edge $p_0p_1$ can be subdivided in $T$. There are
the following two cases:
\begin{packed_itemize}
\item \emph{Case 1.} If $|R'_{v_i}| \ge 3$ for each $i$, i.e, $p_1$ and $p_2$ belong to
each $R'_{v_i}$, the subdivision of $p_0p_1$ is equivalent to the subdivision of $p_1p_2$ which is
correct in the original \sub\ type problem. So nothing needs to be done.
\item \emph{Case 2.} Let $|R'_{v_i}| = 2$ for some $i$, so $R'_{v_i} = \{p_0,p_1\}$. Then $N(v_i)
\setminus v_\leftarrow$ has to form a complete subgraph of $G$, otherwise the starting partial
representation having $R'_{v_i} = \{p_1\}$ would not be extendible. We revert the subdivision of
$p_0p_1$ by modifying $\calR$ as follows. Let $p'_1,\dots,p'_s$ be the new vertices of $T$ created
by the subdivision of $p_0p_1$.  For each $v \in N(v_i) \setminus v_\leftarrow$, we set $R_v = R_v
\setminus \{p'_1,\dots,p'_s\} \cup \{p_1\}$, and we remove $p'_1,\dots,p'_s$ by contractions.
Clearly, the resulting representation is correct and still extends $\calR'$.
\end{packed_itemize}
By removing $p_0$ and the vertices attached to it on the left, $p_{t+1}$ and the vertices attached
to it on the right, $v_\leftarrow$ and $v_\rightarrow$ (of course, only if they are
added), we obtain a correct representation of $G$ inside a subdivision of $T'$ extending the partial
representation $\calR'$.

Concerning \pint, we use almost the same approach. The only difference is that we append two vertices $p_0$
and $\bar p_0$ (resp.~$p_{t+1}$ and $\bar p_{t+1}$) to the end of $T'$, and we put $R'_{v_\leftarrow} =
\{\bar p_0,p_0\}$ (resp.~$R'_{v_\rightarrow} = \{p_{t+1}, \bar p_{t+1}\}$), so the modified partial
representation is proper.\qed
\end{proof}

\subsection{The NP-complete Cases}

The basic gadgets of the reductions are paths. They have the following minimum spans.

\begin{lemma} \label{lem:path}
For\/ \int, $\minspan(P_n) = n$. For\/ \pint\ and $n \ge 2$, $\minspan(P_n) = n+2$.
\end{lemma}

\begin{proof}
For \int, the number of the maximal cliques of $P_n$ is $n$. For \pint, the ordering $\lessdot$ is
$$\ell_0 \lessdot \ell_1 \lessdot r_0 \lessdot \ell_2 \lessdot r_1 \lessdot \cdots \lessdot \ell_i
\lessdot r_{i-1} \lessdot \cdots \lessdot \ell_n \lessdot r_{n-1} \lessdot r_n.$$
There are $n$ changes from $\ell_i$ to $r_{i-1}$ and $P_n$ has $n+1$ vertices. So the minimum span
equals $2(n+1)-n=n+2$.\qed
\end{proof}

We reduce the problems from \partition. An input of \partition\ consists of positive integers $k$, $M$ and
$A_1,\dots,A_{3k}$ such that ${M \over 4} < A_i < {M \over 2}$ for each $A_i$ and $\sum A_i = kM$.
It asks whether it is possible to partition $A_i$'s into $k$ triples such that the sets $A_i$ of each
triple sum to exactly $M$.\footnote{Notice that if a subset of $A_i$'s sums to exactly $M$ it has to
be a triple due to the size constraints.} This problem is strongly \cNP-complete~\cite{partition}
which means that it is \cNP-complete even when the input is coded in unary, i.e., all integers are
of polynomial sizes.

\begin{theorem} \label{thm:int_npc}
The problems $\ext(\pint,\fixed)$ and $\ext(\int,\fixed)$ are \cNP-complete.
\end{theorem}

\begin{proof}
We use almost the same reductions for both \pint\ and \int. For a given input of \partition\ (with
$M \ge 4$), we construct a graph $G$ and its partial representation as follows.

As the fixed tree we choose $T'=P_{(M+1)k}$, with the vertices $p_0,\dots,p_{(M+1)k}$.  The graph
$G$ contains two types of gadgets as separate components. First, it contains $k+1$ \emph{split
gadgets} $S_0,\dots,S_k$ which split the path into $k$ gaps of the size $M$. Then it contains
$3k$ \emph{take gadgets} $T_1,\dots,T_{3k}$. A take gadget $T_i$ takes in each representation at
least $A_i$ vertices of one of the $k$ gaps.

For these reductions, the gadgets are particularly simple. The split gadget $S_i$ is just a single
pre-drawn vertex $v_i$ with $R_{v_i} = \{p_{(M+1)i}\}$. The split gadgets clearly split the path
into the $k$ gaps of the size $M$. The take gadget $T_i$ is $P_{A_i}$ for \int, resp.~$P_{A_i-2}$
for \pint.  According to Lemma~\ref{lem:path}, $\minspan(T_i) = A_i$. The representation is
extendible if and only if it is possible to place the take gadgets into the $k$ gaps. For an
example, see Fig.~\ref{fig:int_partition}. The reduction is clearly polynomial.

\begin{figure}[t]
\centering
\includegraphics{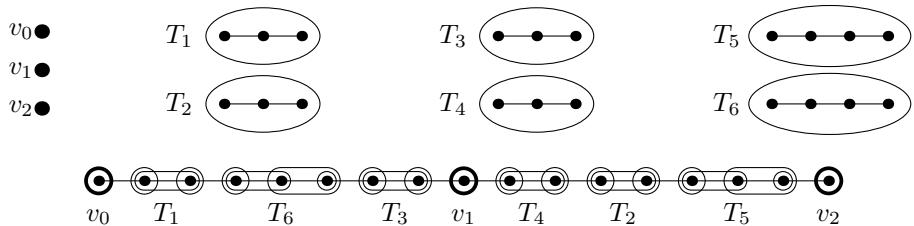}
\caption{An example of the reduction for the following input of \partition: $k = 2$, $M = 7$, $A_1 =
A_2 = A_3 = A_4 = 2$ and $A_5 = A_6 = 3$. On top, the constructed interval graph is depicted. On
bottom, the partial representation (depicted in bold) is extended.}
\label{fig:int_partition}
\end{figure}

To conclude the proof, we show that the partial representation is extendible if and only if the
corresponding \partition\ input has a solution. If the partial representation is extendible, the
take gadgets $T_i$ are divided into the $k$ gaps on the path which gives a partition. Based on the
constraints for the sizes of $A_i$'s, each gap contains exactly three take gadgets of the total minimum
span $M$; thus the partition solves the \partition\ problem. On the other hand, a solution of
\partition\ describes how to place the take gadgets into the $k$ gaps and construct an extending
representation.\qed
\end{proof}

\begin{corollary} \label{cor:int_add_npc}
The problem $\ext(\int,\add)$ is \cNP-complete.
\end{corollary}

\begin{proof}
We use the above reduction for \int\ with one additional pre-drawn interval $v$ attached to
everything in $G$.  We put $R_v = \{p_0,\dots,p_{(M+1)k}\}$, so it contains the whole tree $T'$.
Since a representation of each take gadget $T_i$ has to intersect $R_v$, it has to be placed inside
of the $k$ gaps as before.\qed
\end{proof}

We note that the above modification does not work for proper interval graphs. Indeed, this is not
very surprising since Proposition~\ref{prop:pint_add} states that the problem $\ext(\pint,\add)$ can
be solved in time $\O(n+m)$.

\subsection{The Parameterized Complexity} \label{sec:int_param}

In this subsection, we study the parameterized complexity. The parameters are the number $c$ of components,
the number $k$ of pre-drawn intervals and the size $t$ of the path $T'$.

\heading{By the Number of Components.} In the reduction of Theorem~\ref{thm:int_npc}, one might ask
whether it is possible to make the reduction graph $G$ connected. For \int, it is indeed possible to
add a universal vertex adjacent to everything in $G$, and thus make $G$ connected as in the proof
of Corollary~\ref{cor:int_add_npc}. The following result answers this question for \pint\
negatively (unless $\cP=\cNP$):

\begin{proposition}
The problem $\ext(\pint,\fixed)$ is fixed-parameter trac\-table in the number $c$ of components,
solvable in time $\O((n+m)c!)$.
\end{proposition}

\begin{proof}
There are $c!$ possible orderings $\blt$ of the components from left to right, and we test each of
them.  (The located components force some partial ordering $\blt$ so we need to test less then $c!$
orderings; see below the proof for details.) We show that for a prescribed ordering $\blt$ of the
components, we can solve the problem in time $\O(n+m)$; thus gaining the total time $\O((n+m)c!)$.
We solve the problem almost the same as in the proof of Proposition~\ref{prop:pint_add}. The only
difference is that we deal with all components instead of only the located ones.

We process the components from left to right. When we process $C_i$, we place it on the right of
$C_{i-1}$ as far to the left as possible. For the unlocated $C_i$, we can take any smallest
representation. For the located $C_i$, we test both smallest representations and take the one placing
the right-most endpoint of $C_i$ further to the left.  We construct the representation in time
$\O(n+m)$. For the correctness of the algorithm see the proof of Proposition~\ref{prop:pint_add} for
more details.\qed
\end{proof}

We note that for \cNP-hardness of the problem $\ext(\pint,\fixed)$ it is necessary to have some
pre-drawn subpaths. On the other hand, also some unlocated components are necessary. If all the
components were located, there would be a unique ordering $\blt$ and we could test it in time
$\O(n+m)$ as described above.  In general, for $c$ components and $c'$ located components, we need
to test only $c! \over c'!$ different orderings.

\heading{By the Number of Pre-drawn Intervals.}
In the reduction in Theorem~\ref{thm:int_npc}, we need to have $k$ pre-drawn intervals. One could
ask, whether the problems become simpler with a small number of pre-drawn intervals. We answer this
negatively. For \pint, the problem is in \cXP\ and \cW1-hard with respect to $k$. For \int, we
only show that it is \cW1-hard.

There are two closely related problems \binpack\ and \genbinpack. In both problems, we have
$k$ bins and $n$ items of positive integer sizes. The question is whether we can pack (partition)
these items into the $k$ bins when the volumes of the bins are limited. For \binpack, all the bins have
the same volume. For \genbinpack, the bins have different volumes. Formally:

\computationproblem{\binpack}
{Positive integers $k$, $\ell$, $V$, and $A_1,\dots,A_\ell$.}
{Does there exist a $k$-partition $\calP_1,\dots,\calP_k$ of $A_1,\dots,A_\ell$ such that $\sum_{A_i
\in \calP_j} A_i \le V$ for every $\calP_j$.}

\computationproblem{\genbinpack}
{Positive integers $k$, $\ell$, $V_1,\dots,V_k$, and $A_1,\dots,A_\ell$.}
{Does there exist a $k$-partition $\calP_1,\dots,\calP_k$ of $A_1,\dots,A_\ell$ such that $\sum_{A_i
\in \calP_j} A_i \le V_j$ for every $\calP_j$.}

\begin{lemma}
The problems\/ \binpack\ and\/ \genbinpack\ are polynomially equivalent.
\end{lemma}

\begin{proof}
Obviously \binpack\ is a special case of \genbinpack. On the other hand, let $k$, $\ell$,
$V_1,\dots,V_k$, and $A_1,\dots,A_\ell$ be an instance of \genbinpack. We construct an instance $k'$,
$\ell'$, $V'$, and $A'_1,\dots,A'_{\ell'}$ of \binpack\ as follows. We put $k'=k$, $\ell' = \ell+k$
and $V' = 2\cdot\max V_i + 1$. The sizes of the first $\ell$ items are the same, i.e, $A'_i = A_i$ for
$i=1,\dots,\ell$. The additional items $A'_{\ell+1},\dots,A'_{\ell+k}$ are called \emph{large} and we put
$A'_{\ell+i} = V' - V_i$ for $i=1,\dots,k$.

Each bin has to contain exactly one large item since two large items take more space than $V'$.
After placing large items into the bins, we obtain the bins of the remaining volumes $V_1,\dots,V_k$
in which we have to place the remaining items. This corresponds exactly to the original \genbinpack\
instance.\qed
\end{proof}

If the sizes of items are encoded in binary, the problem is \cNP-complete even for $k = 2$. The more
interesting version which we use here is that the sizes are encoded in unary so all sizes are
polynomial. In such a case, the \binpack\ problem is known to be solvable in time $t^{\O(k)}$ using
dynamic programming where $t$ is the total size of all items. And it is \cW1-hard with respect to
the parameter $k$~\cite{binpacking}. The similar holds for $\ext(\pint,\fixed)$:

\begin{proof}[Theorem~\ref{thm:pint_equiv_bin_pack}]
For a given instance of the \binpack\ problem, we can solve it by $\ext(\pint,\fixed)$ in a similar
manner as in the reduction in Theorem~\ref{thm:int_npc}. As $T'$, take a path $P_{(V+1)k}$. As $G$,
take $P_{A_i-2}$ for each $A_i$ and the pre-drawn vertices $v_0,\dots,v_k$ such that $R_{v_i} =
\{p_{(V+1)i}\}$. The rest of the argument is exactly as in the proof of Theorem~\ref{thm:int_npc}.

Now, we want to solve $\ext(\pint,\fixed)$ using $2^k$ instances of \genbinpack\ (which is
polynomially equivalent to \binpack), where $k$ is the number of pre-drawn intervals.

First we deal with located components $C_1 \blt \cdots \blt C_c$. For each component, we have two possible
orderings $\wlt$ and using Lemma~\ref{lem:pint_minspan} we get (at most) two possible smallest
representations which might be differently shifted. In total, we have at most $2^c \le 2^k$ possible
representations keeping $C_1,\dots,C_c$ as small as possible leaving maximal gaps for unlocated
components. We test each of these $2^c$ representations.

Let $C'_1,\dots,C'_{c'}$ be the unlocated components. For each $C'_i$, we compute $\minspan(C'_i)$
using Lemma~\ref{lem:pint_minspan}. The goal is to place the unlocated components into the $c+1$
gaps between representations of the located components $C_1,\dots,C_c$. We can solve this problem
using \genbinpack\ as follows. We have $k+1$ bins of the volumes equal to the sizes of the gaps
between the representations of $C_1,\dots,C_c$. We have $c'$ items of the sizes $A_i =
\minspan(C'_i)$.

A solution of \genbinpack\ tells how to place the unlocated components into the $k$ gaps.
If there exists no solution, this specific representation of the located components cannot
be used. We can test all $2^c$ possible representations of the located components. Thus we get the
required weak truth-table reduction.\qed
\end{proof}

\begin{corollary} \label{cor:pint_param_k}
The problem $\ext(\pint,\fixed)$ is \cW1-hard and belongs to \cXP, solvable in time $n^{\O(k)}$ where
$k$ is the number of pre-drawn intervals.
\end{corollary}

\begin{proof}
Both claims follow from Theorem~\ref{thm:pint_equiv_bin_pack}.\qed
\end{proof}

\begin{proposition} \label{prop:ext_int_hard_in_k}
The problems $\ext(\int,\fixed)$ and $\ext(\int,\add)$ are \cW1-hard when parameterized by the
number $k$ of pre-drawn intervals.
\end{proposition}

\begin{proof}
We modify the reductions of Theorem~\ref{thm:int_npc} and Corollary~\ref{cor:int_add_npc} exactly as
in the proof of Theorem~\ref{thm:pint_equiv_bin_pack}.\qed
\end{proof}

\heading{By the Size of the Path.}
We show that the \fixed\ type problems are fixed-parameter tractable with respect to the size of the
path $t$. It is easy to find a solution by a brute-force algorithm:

\begin{proposition} \label{prop:int_fixed_size_fpt}
For the size $t$ of a path $T'$, the problems $\ext(\pint,\fixed)$ and $\ext(\int,\fixed)$ are
fixed-parameter tractable with the respect to the parameter $t$. They can be solved in time
$\O(n+m+f(t))$ where
$$f(t) = t^{2t^2}.$$
\end{proposition}

\begin{proof}
In a pruned graph, the vertices have to be represented by pairwise different intervals. There are at
most $t^2$ possible different subpaths of a path with $t$ vertices so the pruned graph can contain
at most $t^2$ vertices; otherwise the extension is clearly not possible. We can test every possible
assignment of the non-pre-drawn vertices to the $t^2$ subpaths, and for each assignment we test
whether we get a correct representation extending $\calR'$.\qed
\end{proof}

\section{Path and Chordal Graphs} \label{sec:chor}

We present and prove the results concerning the classes \path\ and \chor.

\subsection{The Polynomial Cases}

The recognition problems for the types \add\ and \both\ are equivalent to standard recognition without any
specified tree $T'$. Indeed, we can modify $T'$ by adding an arbitrary tree to it. If the input
graph is \path\ or \chor, there exists a tree $T''$ in which the graph can be represented. We
produce $T$ by attaching $T''$ to $T'$ in any way. Then the input graph can be represented in $T$ as
well, completely ignoring the part $T'$.

For path graphs, the original recognition algorithm is due to Gavril~\cite{path_original} in time
$\O(n^4)$. The current fastest algorithm is by Sch\"affer~\cite{path_fastest} in time $\O(nm)$. For
chordal graphs, there is a beautiful simple algorithm by Rose et al.~\cite{recog_chordal_graphs} in
time $\O(n+m)$.

\subsection{The NP-complete Cases}

All the remaining cases from the table of Fig.~\ref{fig:complexity_table} are \cNP-complete.  We
modify the reduction for \int\ of Theorem~\ref{thm:int_npc}. We start with the simplest reduction
for the \fixed\ type and then modify it for the other types.

\heading{Fixed Type Recognition.}
For the \fixed\ type, we can avoid pre-drawn subtrees, using an additional structure of the tree.

\begin{proposition} \label{prop:path_fixed_npc}
The problems\/ $\recog^*(\path,\fixed)$ and\/ $\recog^*(\chor,\fixed)$ are \cNP-complete.
\end{proposition}

\begin{proof}
We again reduce from \partition\ with an input $k$ and $M$. For technical purposes, let $M \ge 8$
and so $|A_i| > 2$ for each $A_i$.  We construct a graph $G$ and a tree $T'$ as follows.

The tree $T'$ is a path $P_{(M+1)k}$ (its vertices being denoted by $p_0,\dots,p_{(M+1)k}$) with three paths
of length two attached to every vertex $p_{(M+1)i}$, for each $i = 0,\dots,k$; see
Fig.~\ref{fig:path_partition}.  Each split gadget $S_i$ is a star, depicted on the left of
Fig.~\ref{fig:path_partition}.  When the split gadgets are placed as in $T'$, they split the tree
into $k$ gaps exactly as the pre-drawn vertices in the proof of Theorem~\ref{thm:int_npc}. Each
take gadget $T_i$ is the path $P_{A_i}$ exactly as before. The reduction is obviously polynomial.

\begin{figure}[t]
\centering
\includegraphics{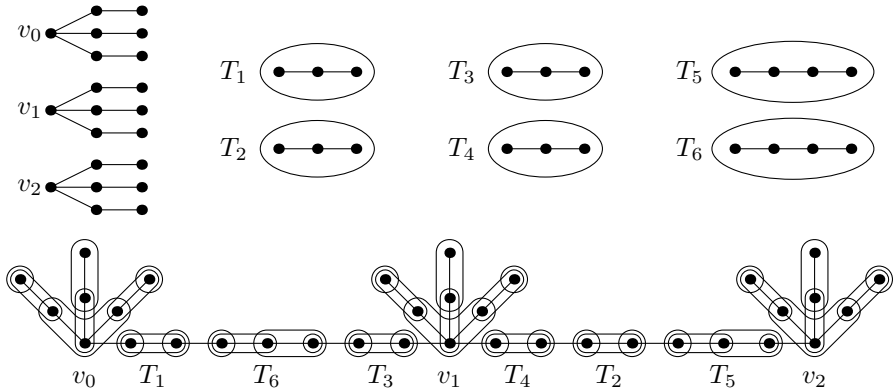}
\caption{An example for the same input of \partition\ as in Fig.~\ref{fig:int_partition}. On top the
graph $G$ is depicted.  On bottom, a representation of $G$ is constructed, giving the solution
$\{A_1,A_3,A_6\}$ and $\{A_2,A_4,A_5\}$.}
\label{fig:path_partition}
\end{figure}

What remains to argue is the correctness of the reduction. Observe that given a solution of
\partition, we can construct a subpaths-in-tree representation of $G$ as in
Fig.~\ref{fig:path_partition}. For the other direction, let $v_0,\dots,v_k$ be the central vertices
of the split gadgets $S_0,\dots,S_k$. We claim that each $R_{v_i}$ contains at least one branch
vertex. (Actually, exactly one since there are only $n+1$ branch vertices in $T'$.) If some
$R_{v_i}$ contained only non-branch vertices, then it would not be possible to represent three
disjoint neighbors $u_1$, $u_2$ and $u_3$ of this $v_i$ having each $R_{u_j} \setminus R_{v_i}$
non-empty.

Since each branch vertex is taken by one $R_{v_i}$, the path $P_{(M+1)k}$ is split into $k$ gaps as
before. Since $|A_i|>2$, each $T_i$ can be represented only inside of these gaps. Notice that the
total number of the vertices in the gaps has to be equal $kM$, and therefore the split gadgets have
to be represented entirely in the attached stars as in Fig.~\ref{fig:path_partition}. The rest of
the reduction works exactly as in Theorem~\ref{thm:int_npc}.\qed
\end{proof}

\heading{Sub Type Recognition.}
By modifying the above reduction, we get:

\begin{theorem} \label{thm:path_sub_npc}
The problems $\recog^*(\path,\sub)$ and $\recog^*(\chor,\sub)$ are \cNP-complete.
\end{theorem}

\begin{proof}
We need to modify the two gadgets from the reduction of Theorem~\ref{prop:path_fixed_npc} in such a
way that a subdivision of the tree $T'$ does not help in placing them. Subdivision only increases the
number of non-branch vertices. Thus a take gadget $T_i$ requires $A_i$ branch vertices.
Similarly, the split gadget $S_i$ is more complicated. See Fig.~\ref{fig:path_sub_partition} on top.

The tree $T$ is constructed as follows. We start with a path $P_{(M+1)k}$ with vertices
$p_0,\dots,p_{(M+1)k}$. To each vertex $p_{(M+1)i}$ we attach a subtree isomorphic to the trees in
Fig.~\ref{fig:path_sub_partition} on bottom. To the remaining vertices of the path, we attach one
leaf per vertex. The reduction is again polynomial.

Straightforwardly, for a given solution of \partition, we can construct a correct subpaths-in-tree representation
in a subdivided tree. On the other hand, we are going to show how to construct a solution of
\partition\ from a given tree representation.

Recall the properties of maximal cliques from Section~\ref{sec:prelim}. Note that each triangle $u_1u_2u_3$ in each
split or take gadget is a maximal clique $K$. Since $N[u_i] \ne N[u_j]$ for each $i \ne j$, there
has to be a branch vertex in $R_{u_i} \cap R_{u_j}$ for some $i$ and $j$. The gadget $S_i$ contains
three triangles, each taking one branch vertex of $T$. In addition, $R_{v_i}$ connecting them has to
contain another branch vertex. So in total, $S_i$ contains at least four branch vertices. Each
gadget $T_i$ contains $A_i$ triangles, and so it requires at least $A_i$ branch vertices. Since the
number of branch vertices of $T$ is limited, each $S_i$ takes exactly four branch vertices and each
$T_i$ takes exactly $A_i$ branch vertices.

\begin{figure}[t]
\centering
\includegraphics{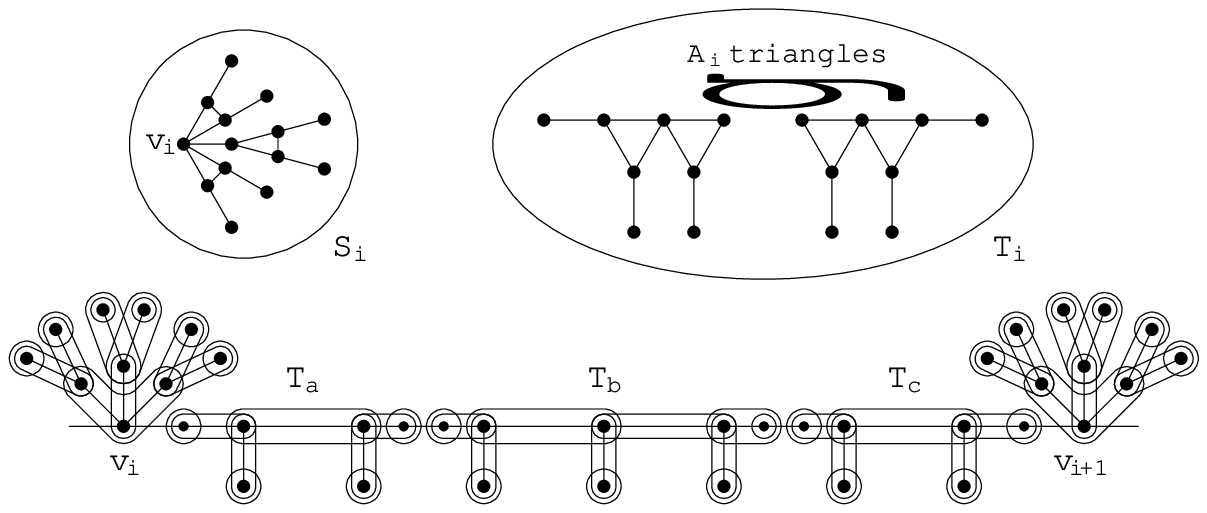}
\caption{On top, the split gadget $S_i$ is on the left and the take gadget $T_i$ is on the right. On
bottom, a part of the tree $T$ is depicted with the small vertices added by subdivisions. The gap between two
split gadgets contains three take gadgets $T_a$, $T_b$ and $T_c$ giving one triple $\{A_a,A_b,A_c\}$
with $A_a + A_b + A_c = M$.}
\label{fig:path_sub_partition}
\end{figure}

Now, if some $T_i$ contained a branch vertex of the subtrees attached to $p_{(M+1)j}$, at least
one of its branch vertices would not be used. (Either not taken by $T_i$, or $T_i$ would require at least
$A_i + 1$ branch vertices.) So each $S_i$ has to take the branch vertices of the subtrees attached
to $p_{(M+1)j}$ for some $j$, and the take gadgets have to be placed inside the gaps exactly as
before.\qed
\end{proof}

\begin{proposition} \label{prop:chor_add_both_npc}
Even with a single pre-drawn subtree, i.e, $|G'| = 1$, the problems $\ext(\chor,\add)$ and
$\ext(\chor,\both)$ are \cNP-complete.
\end{proposition}

\begin{proof}
We easily modify the above reductions; for the \add\ type, the reduction of
Proposition~\ref{prop:path_fixed_npc}, for the \both\ type, the reduction of
Theorem~\ref{thm:path_sub_npc}.  The modification adds into $G$ one pre-drawn vertex $v$ adjacent to
everything such that $R_v = T'$.  Since $R_v$ spans the whole tree, it forces the entire
representation $\calR$ into $T'$.

We just deal with the \add\ type, for \both\ the argument is exactly the same. Let $T'$ be the
partial tree and let $T$ be the tree in which the representation is constructed, so $T'$ is a
subtree of $T$. We claim that we can restrict a representation of each vertex of $G$ into $T'$ and
thus obtain a correct representation inside the subtree $T'$.

Let $x \in V$. Since $xv \in E(G)$, the intersection of $R_x$ and $T'$ is a non-empty subtree. We put
$\widetilde R_x = R_x \cap T'$, and we claim that $\widetilde\calR$ is a representation of $G$ in
$T'$.  To argue the correctness, let $x$ and $y$ be two different vertices from $v$ (otherwise
trivial). If $xy \notin E(G)$, then $R_x \cap R_y = \emptyset$, and so $\widetilde R_x \cap \widetilde
R_y = \emptyset$ as well. Otherwise $xyv$ is a triangle in $G$, and thus by the Helly property the
subtrees $R_x$, $R_y$ and $R_v = T'$ have a non-empty common intersection, giving that $\widetilde
R_x \cap \widetilde R_y$ is non-empty.\qed
\end{proof}

For path graphs, one can use a similar technique of a pre-drawn universal vertex attached to
everything. But there is the following difficulty: To do so, the input partial tree $T'$ has to be
a path. For the type \both, the complexity of $\ext(\path,\both)$ remains open. For the type \add, we get the
following weaker result:

\begin{proposition}
The problem $\ext(\path,\add)$ is \cNP-complete.
\end{proposition}

\begin{proof}
Similarly as in Proposition~\ref{prop:chor_add_both_npc}, add a pre-drawn universal vertex $v$ on the
path $T'$ constructed in the reduction of Theorem~\ref{thm:int_npc} such that $R_v = T'$. The rest
is exactly as above.\qed
\end{proof}

\subsection{The Parameterized Complexity}

We deal with parameterized complexity of the problems and we give only minor and partial results in
this direction. Unlike in Section~\ref{sec:int}, parameterization by the number $k$ of pre-drawn
subtrees is mostly not helpful. We show that every problem with exception of $\ext(\path,\add)$ is
already \cNP-complete for $k=0$ or $k=1$. For $\ext(\path,\add)$, we have only a weaker result that
it is \cW1-hard with respect to the parameter $k$ since Proposition~\ref{prop:ext_int_hard_in_k}
straightforwardly generalizes.

Similarly, a low number $c$ of components does not make the problem any easier. We can easily
insert a universal vertex attached to everything. So the above reductions can be modified and the
problems remain \cNP-complete even if the graph $G$ is connected.

Concerning the size $t$ of the tree, Proposition~\ref{prop:int_fixed_size_fpt} straightforwardly
generalizes:

\begin{proposition} \label{thm:hfixed}
Let $t$ be the size of $T'$. The problems $\ext(\path,\fixed)$ and $\ext(\chor,\fixed)$ are
fixed-parameter tractable with respect to $t$. They can be solved in time $\O(n+m+g(t))$ where
$$g(t) = 2^{t2^t}.$$
\end{proposition}

\begin{proof}
Proceed exactly as in the proof of Proposition~\ref{prop:int_fixed_size_fpt}, test all possible
assignments of all vertices of a pruned graph. The only difference is that $T'$ has at most $2^t$
different subtrees.\qed
\end{proof}

We note that a more precise bound for the number of subtrees could be use but we did not try to
better estimate the function $g$.

\section{Conclusions} \label{sec:conclusions}

In this paper, we have considered different problems concerning extending partial representations of
chordal graphs and their three subclasses. One of the main goals of this paper is to stimulate
a future research in this area. Therefore, we conclude with three open problems.

The first problem concerns the only open case in the table in Fig.~\ref{fig:complexity_table}.

\begin{problem}
What is the complexity of $\ext(\path,\both)$?
\end{problem}

Concerning the parameterized complexity, we believe it is useful to first attack problems related to
interval graphs. This allows to develop tools for more complicated chordal graphs. A generalization
of Theorem~\ref{thm:pint_equiv_bin_pack} and Corollary~\ref{cor:pint_param_k} for \int\ seems to be
particularly interesting. The PQ-tree approach seems to be a good starting point.

\begin{problem}
Does $\ext(\int,\fixed)$ belong to \cXP\ with respect to $k$ where $k$ is the number of pre-drawn
intervals?
\end{problem}

We present only basic results concerning the parameterized complexity with respect to the parameter
$t$ where $t$ is the size of the tree $T'$. We deal with the type \fixed\ for which the solution is
straightforward. The complexity for the other types \sub, \add, and \both\ remains open.

\begin{problem}
What is the parameterized complexity of the remaining problems with respect to the parameter $t$
where $t$ is the size of the tree $T'$?
\end{problem}

\begin{acknowledgements}
The first two authors are supported by ESF Eurogiga project GraDR as GA\v{C}R GIG/11/E023. The first
author is also supported by Charles University as GAUK 196213.
\end{acknowledgements}

\bibliographystyle{spmpsci}
\bibliography{extending_chor_journal}

\begin{thebibliography}{10}
\providecommand{\url}[1]{{#1}}
\providecommand{\urlprefix}{URL }
\expandafter\ifx\csname urlstyle\endcsname\relax
  \providecommand{\doi}[1]{DOI~\discretionary{}{}{}#1}\else
  \providecommand{\doi}{DOI~\discretionary{}{}{}\begingroup
  \urlstyle{rm}\Url}\fi

\bibitem{blas_rutter}
Bl{\"a}sius, T., Rutter, I.: Simultaneous {PQ}-ordering with applications to
  constrained embedding problems.
\newblock In: SODA '13 (2013)

\bibitem{PQ_trees}
Booth, K.S., Lueker, G.S.: Testing for the consecutive ones property, interval
  graphs, and planarity using {PQ}-tree algorithms.
\newblock Journal of Computational Systems Science \textbf{13}, 335--379 (1976)

\bibitem{cfk}
Chaplick, S., Fulek, R., Klav\'{\i}k, P.: Extending partial representations of
  circle graphs.
\newblock In preparation.  (2013)

\bibitem{uint_corneil}
Corneil, D.G., Kim, H., Natarajan, S., Olariu, S., Sprague, A.P.: Simple linear
  time recognition of unit interval graphs.
\newblock Information Processing Letters \textbf{55}(2), 99--104 (1995)

\bibitem{LBFS_int}
Corneil, D.G., Olariu, S., Stewart, L.: The {LBFS} structure and recognition of
  interval graphs.
\newblock SIAM Journal on Discrete Mathematics \textbf{23}(4), 1905--1953
  (2009)

\bibitem{deng}
Deng, X., Hell, P., Huang, J.: Linear-time representation algorithms for proper
  circular-arc graphs and proper interval graphs.
\newblock SIAM J. Comput. \textbf{25}(2), 390--403 (1996)

\bibitem{fiala}
Fiala, J.: {NP} completeness of the edge precoloring extension problem on
  bipartite graphs.
\newblock J. Graph Theory \textbf{43}(2), 156--160 (2003)

\bibitem{maximal_cliques}
Fulkerson, D.R., Gross, O.A.: Incidence matrices and interval graphs.
\newblock Pac. J. Math. \textbf{15}, 835--855 (1965)

\bibitem{partition}
Garey, M.R., Johnson, D.S.: Complexity results for multiprocessor scheduling
  under resource constraints.
\newblock SIAM Journal on Computing \textbf{4}(4), 397--411 (1975)

\bibitem{chor_is_subtree_in_tree}
Gavril, F.: The intersection graphs of subtrees in trees are exactly the
  chordal graphs.
\newblock Journal of Combinatorial Theory, Series B \textbf{16}(1), 47--56
  (1974)

\bibitem{path_original}
Gavril, F.: A recognition algorithm for the intersection of graphs of paths in
  trees.
\newblock Discrete Mathematics \textbf{23}, 211--227 (1978)

\bibitem{agt}
Golumbic, M.C.: Algorithmic Graph Theory and Perfect Graphs.
\newblock North-Holland Publishing Co. (2004)

\bibitem{prext_bip}
Hujter, M., Tuza, Z.: Precoloring extension. {II.} {G}raph classes related to
  bipartite graphs.
\newblock Acta Mathematica Universitatis Comenianae \textbf{62}(1), 1--11
  (1993)

\bibitem{simrep_chor}
Jampani, K., Lubiw, A.: The simultaneous representation problem for chordal,
  comparability and permutation graphs.
\newblock In: Algorithms and Data Structures, \emph{Lecture Notes in Computer
  Science}, vol. 5664, pp. 387--398 (2009)

\bibitem{binpacking}
Jansen, K., Kratsch, S., Marx, D., Schlotter, I.: Bin packing with fixed number
  of bins revisited.
\newblock In: Algorithm Theory - SWAT 2010, \emph{Lecture Notes in Computer
  Science}, vol. 6139, pp. 260--272 (2010)

\bibitem{kkkw}
Klav{\'i}k, P., Kratochv{\'i}l, J., Krawczyk, T., Walczak, B.: Extending
  partial representations of function graphs and permutation graphs.
\newblock In: Algorithms -- ESA 2012, \emph{Lecture Notes in Computer Science},
  vol. 7501, pp. 671--682 (2012)

\bibitem{kkorssv}
Klav\'{\i}k, P., Kratochv\'{\i}l, J., Otachi, Y., Rutter, I., Saitoh, T.,
  Saumell, M., Vysko\v{c}il, T.: Extending partial representations of proper
  and unit interval graphs.
\newblock In preparation.  (2012)

\bibitem{kkos}
Klav\'{\i}k, P., Kratochv\'{\i}l, J., Otachi, Y., Saitoh, T.: Extending partial
  representations of subclasses of chordal graphs.
\newblock In: Algorithms and Computation -- ISAAC, \emph{Lecture Notes in
  Computer Science}, vol. 7676, pp. 444--454 (2012)

\bibitem{kkosv}
Klav\'{\i}k, P., Kratochv\'{\i}l, J., Otachi, Y., Saitoh, T., Vysko\v{c}il, T.:
  Linear-time algorithm for partial representation extension of interval
  graphs.
\newblock Journal version of the TAMC paper, in preparation.  (2012)

\bibitem{kkv}
Klav\'{\i}k, P., Kratochv\'{\i}l, J., Vysko\v{c}il, T.: Extending partial
  representations of interval graphs.
\newblock In: Theory and Applications of Models of Computation - 8th Annual
  Conference, TAMC 2011, \emph{Lecture Notes in Computer Science}, vol. 6648,
  pp. 276--285 (2011)

\bibitem{proper_first}
Looges, P.J., Olariu, S.: Optimal greedy algorithms for indifference graphs.
\newblock Comput. Math. Appl. \textbf{25}, 15--25 (1993)

\bibitem{tig}
McKee, T.A., McMorris, F.R.: Topics in Intersection Graph Theory.
\newblock SIAM Monographs on Discrete Mathematics and Applications (1999)

\bibitem{recog_chordal_graphs}
Rose, D.J., Tarjan, R.E., Lueker, G.S.: Algorithmic aspects of vertex
  elimination on graphs.
\newblock SIAM Journal on Computing \textbf{5}(2), 266--283 (1976)

\bibitem{path_fastest}
Sch\"affer, A.A.: A faster algorithm to recognize undirected path graphs.
\newblock Discrete Appl. Math \textbf{43}, 261--295 (1993)

\bibitem{egr}
Spinrad, J.P.: Efficient Graph Representations.
\newblock Field Institute Monographs (2003)

\end{thebibliography}

\end{document}